\documentstyle[twocolumn]{mn}
\newcommand{\mincir}{\raise -2.truept\hbox{\rlap{\hbox{$\sim$}}\raise5.truept
\hbox{$<$}\ }}
\newcommand{\magcir}{\raise -2.truept\hbox{\rlap{\hbox{$\sim$}}\raise5.truept
\hbox{$>$}\ }}
\newcommand{\minmag}{\raise-2.truept\hbox{\rlap{\hbox{$<$}}\raise 6.truept\hbox
{$>$}\ }}
\newcommand{\be}{\begin{equation}}
\newcommand{\ee}{\end{equation}}
\newcommand{\ba}{\begin{eqnarray}}
\newcommand{\ea}{\end{eqnarray}}
\newcommand{\brr}{\begin{array}}
\newcommand{\err}{\end{array}}
\newcommand{\bc}{\begin{center}}
\newcommand{\ec}{\end{center}}

\newcommand{\bv}{\mbox{\bf v}}

\newcommand{\bq}{\mbox{\bf q}}
\newcommand{\bx}{\mbox{\bf x}}

\newcommand{\lb}{{\left<\right.}}
\newcommand{\rb}{{\left.\right>}}
\newcommand{\hm}{\,h^{-1}{\rm Mpc}}
\newcommand{\vel}{\,{\rm km\,s^{-1}}}

\newcommand{\xvec}{{\bf x}}

\title[The Cluster Velocity Field]{ The Cluster Distribution as a Test of
Dark Matter Models. III: The Cluster Velocity Field}
\author[L. Moscardini et al.]
{ L. Moscardini$^1$, E. Branchini$^{2,3}$,
P. Tini Brunozzi$^{4,5}$, S. Borgani$^{2,5}$, \\
{\LARGE M. Plionis$^{2,6}$ \& P. Coles$^7$} \\
$^1$Dipartimento di Astronomia, Universit\`a di Padova,
vicolo dell'Osservatorio 5, I--35122 Padova, Italy  \\
$^2$SISSA -- International School for Advanced Studies,
via Beirut 2--4, I--34013 Trieste, Italy \\
$^3$Department of Physics, University of Durham, South Road, Durham DH1
3LE, UK \\
$^4$Dipartimento di Fisica, Universit\`a di Perugia,
via A. Pascoli, I--06100 Perugia, Italy \\
$^5$INFN Sezione di Perugia, c/o Dipartimento di Fisica dell'Universit\`{a},
via A. Pascoli, I--06100 Perugia, Italy \\
$^6$National Observatory of Athens, Lofos Nimfon, Thesio, 18110 Athens,
Greece \\
$^7$Astronomy Unit, School of Mathematical Sciences,
Queen Mary \& Westfield College, Mile End Road, London E1 4NS, UK }
\begin{document}

\maketitle

\begin{abstract}
We study the large--scale velocity fields traced by galaxy clusters in
numerical simulations of a box of side 960 $h^{-1}$ Mpc, and compare them with
available data on real clusters. In order to test the reliability of the
simulations, which are based on an optimized version of the Zel'dovich
approximation, we compare their cluster velocities  with those of ``exact"
N--body simulations, and find a remarkable agreement between the two according
to a variety of  statistical tests. We analyse Cold Dark Matter (CDM) models
with density parameter in the range $0.2\le \Omega_0\le 1$, both with and
without the cosmological constant term to provide a flat geometry. We also
simulate a Cold+Hot Dark Matter (CHDM) model, with 30\% provided by the hot
component. Comparison with real data is performed by applying tests based on
the cumulative velocity frequency distribution (CVFD) and bulk flow statistics.
For the CVFD, we use observational velocity data from different authors. By
merging all available data in a combined sample, the CVFD test is able to
exclude only open models with $\Omega_0\le 0.4$ and flat models with
$\Omega_0=0.2$. However, the analysis of individual observational samples gives
contradictory results; in particular the recent Giovanelli (1995) data, which
imply much lower velocities (with significantly smaller uncertainties) than the
other data, are actually only consistent with the open $\Omega_0=0.4$ and the
flat $\Omega_0=0.2$ models, i.e. the models which are excluded by the other
data. Furthermore, CVFD and bulk flow analyses of the Branchini et al. (1995)
reconstructed velocity data disfavour precisely those models accepted on the
grounds of Giovanelli's sample. Finally, we confirm that the Lauer \& Postman
(1994) bulk flow determination is an extremely rare event in the cosmological
models we have analysed.
\end{abstract}

\begin{keywords}
Cosmology: theory -- dark matter -- galaxies: clustering, formation --
large--scale structure of Universe
\end{keywords}

$$
{}~~~
$$
\newpage
\section{Introduction}

The quality and quantity of galaxy peculiar velocity data
derived from the Infra--Red
Tully--Fisher (IRTF) and $D_n$--$\sigma$ scaling relations is progressively
increasing,  providing an ever more precise description of
large--scale cosmic flows. Moreover, an independent technique to estimate the
distance of high redshift galaxies with very small errors, based on the shapes
of type Ia supernova light curves, recently developed by Riess,
Press \& Kirshner (1995), could allow a description of
motions on even larger scales. Consequently the study of large--scale
flows is potentially one of the most powerful tools with which to
 understand the structure of the universe on large ($\magcir 50\,h^{-1}$ Mpc)
scales (see reviews by Dekel 1994; Strauss \& Willick 1995 and
references therein).
Furthermore, the
hypothesis of gravitational instability allows one to relate the
peculiar velocity field directly
to the entire mass distribution, under the plausible
assumption that at such large scales galaxies share the same dynamics as
the dark matter distribution.

Clusters of galaxies provide a particularly efficient and precise way
to estimate peculiar motions, since the determination of many
redshift--independent distances of cluster members leads to
a substantial reduction in the statistical errors compared to
studies of isolated galaxies.
Moreover, new techniques based on the Sunyaev--Zel'dovich (1980) effect
(Haenhelt \& Tegmark 1995) and on the brightest cluster galaxy distance
indicator method (Lauer \& Postman 1994) are beginning to provide independent
estimates of cluster peculiar velocities whose reliability is expected to
increase considerably in the next few years.

A further important motivation for considering clusters
is that they can sample the cosmic velocity field up to distances larger
than those presently accessible by galaxies.
On these larger scales, fluctuations are largely in the linear
regime and are, therefore, more
closely related to the ``initial conditions'' from which large--scale
structure developed. In addition,
since large--scale velocities are mostly sensitive to long--wavelength
density fluctuations, one expects that, for a fixed choice of the
fundamental cosmological parameters (Hubble constant $H_0$, density
parameter $\Omega_0$ and cosmological constant term $\Omega_\Lambda$),
the amplitude of large--scale motions is almost independent of the composition
of the dark
matter (DM) component (i.e., cold vs. hot DM). For these reasons, several
authors (e.g., Bahcall, Cen \& Gramann 1994; Croft \& Efstathiou 1994;
Bahcall, Gramann \& Cen 1994; Cen, Bahcall \& Gramann 1994; Strauss et al.
1995; Gramann et al. 1995) have  recently addressed the issue
of whether  cluster motions can be used to constrain cosmological models.
Various statistical tests have been used for this purpose,
among which the velocity
frequency distribution, bulk flows, velocity dispersions, pairwise
velocities and the velocity correlation function feature prominently.

The present work is devoted to a quantitative comparison between presently
available observational data on cluster peculiar velocities and an
extended set of numerical simulations of a number of different
cosmological models. The simulations are based on an optimized version of
the Zel'dovich approximation (Zel'dovich 1970; Shandarin \&
Zel'dovich 1989; Coles, Melott \& Shandarin 1993),
the ability of which to simulate the large--scale
distribution of galaxy clusters has been assessed by Borgani et al. (1995;
see also Plionis et al. 1995).
We will show in the following that the quasi--linear description of the
large--scale velocity field provided by our simulation method reproduces to a
good accuracy that obtained from N--body simulations. At the same time, the
method is so cheap, computationally speaking,
that it enables us to consider easily a large
number of realizations of a large number of different scenarios.

In this paper we consider $\Omega_0<1$ CDM models
both with and without a cosmological constant term $\Lambda$ imposing the
condition of spatial flatness: $\Omega_\Lambda=\Lambda/(3 H_0^2)$.
We also consider, as reference models, the $\Omega_0=1$ CDM and
CHDM scenarios. Our main aim here is to investigate whether the
cluster velocity field can constrain the values of $\Omega_0$ and
$\Omega_\Lambda$. (In fact, for a fixed spectrum normalization, the
resulting motions essentially depend on the value of $\Omega_0$: the
larger the density parameter, the larger are the peculiar velocities.)

Observational motivations for cosmological scenarios with $\Omega_0 <1$
have been considered by different authors (cf. Coles \& Ellis 1994; see
Ratra \& Peebles 1995 for a summary; see, however, Primack 1995, for a
different view about a low--$\Omega_0$ Universe). Due to the fact that
``ordinary'' inflation models can accommodate open universes only by fine
tuning
either the duration of inflation or the pre-inflationary conditions (e.g.
Lucchin \& Matarrese 1985; Ellis, Lyth \& Mijic 1991),
interest, in the past, has been
limited only to models in which $\Omega_\Lambda \neq 0$ (Efstathiou,
Sutherland \& Maddox 1990; Kofman, Gnedin \& Bahcall
1993). Only very recently has it been realized that, in the framework of the
bubble nucleation model, it is possible to obtain an open geometry,
regardless of the initial conditions, but requiring only an
appropriate choice for the scalar field potential (Sasaki et al. 1993;
Bucher, Goldhaber \& Turok 1995); see
Linde \& Mezhlumian (1995), and references therein, for a review of this
class of inflationary models.  An extensive comparison of the predictions
of open CDM models with a variety of observational data is given by
Liddle et al. (1995) and Yamamoto \& Bunn (1995); see also Ratra \& Peebles
(1994).

The layout of this paper is as follows. In Section 2 we present our approach
for simulating the cluster velocity field and we discuss its reliability by
comparing relevant results with those obtained using an N--body code.
We also describe the statistical
tests applied in the analysis and briefly introduce the suite of
 cosmological models we consider.
 In Section 3 we compare the output for different initial spectra
in terms of both the velocity frequency distribution and the velocity
correlation function. Section 4 is devoted to a comparison with
available observational
data; in particular we study the bulk flow and the cumulative velocity
frequency distribution. In Section 5 we discuss our results and, finally,
we draw our main conclusions in Section 6.

\section{The simulations}
\subsection{The Zel'dovich approach}
Our simulations mimic  gravitational dynamics through the Zel'dovich
approximation (ZA; Zel'dovich 1970; Shandarin \& Zel'dovich 1989). This method
is based on the Eulerian--to--Lagrangian space mapping
\be
\bx(\bq,t)\,=\,\bq-b(t){\bf \nabla}\psi(\bq)\,,
\label{eq:za}
\ee
where $\bq$ and $\bx$ are initial and final particle positions, respectively,
$b(t)$ is the fluctuation linear growth factor and the velocity potential
$\psi(\bq)$ is related to the initial fluctuation field by the Poisson
equation. The simulation procedure has been described in detail in Borgani et
al. (1995; hereafter Paper I), to which we refer the reader
for more details. Here we simply recall
that the initial power spectrum is convolved with a Gaussian window,
$P(k)\to P(k)e^{-k^2R_f^2}$, in order to suppress shell--crossing at small
scales (Coles, Melott \& Shandarin 1993; Melott, Pellman \& Shandarin
1994). The filtering radius $R_f$ is chosen in such a way that the average
number of streams at each Eulerian point (Kofman et al. 1994) is $N_s=1. 1$.
After assigning the linear potential on a grid, particles are moved from their
initial grid positions according to eq.(\ref{eq:za}). The density and potential
fields are then reassigned on the grid and clusters are selected as the highest
local density maxima, so as to reproduce the observed Abell/ACO cluster number
density, $n_{cl}\simeq 1.8\times 10^{-5}(\hm)^{-3}$, which corresponds to
$d_{cl}\simeq 38\hm$ for the mean cluster separation. The parameter $h$ is the
Hubble constant in units of $100 \vel$ Mpc$^{-1}$.

Velocities at the grid positions are evaluated by following two different
prescriptions.
\begin{description}
\item[(a)] Defining the particle velocities as $\bv\propto -{\bf
\nabla}\psi(\bq)$ and interpolating their mass and momentum on the grid
with the TSC scheme (Hockney \& Eastwood 1981).
The velocity at the grid point is then defined as the ratio between the local
values of momentum and mass (e.g., Kofman et al. 1994). This linear
prescription, that we will name LIN in the following, amounts to
the assumption that
particles at the final positions do not still feel any tidal force
and move according to the initial ({\em linear}) gravitational potential.
\item[(b)] Computing the gradient of the ZA--evolved potential,
$\psi_{_{ZA}}$,
which is connected by the Poisson equation to the density fluctuations,
$\delta_{_{ZA}}$, traced by the final particle distribution,
$\nabla^2\psi_{_{ZA}}=4\pi G\bar \rho a^2\delta_{_{ZA}}$.
Accordingly, $\bv \propto
-{\bf \nabla}\psi_{_{ZA}}$.
In this quasi--linear prescription, named ZEL in the
following, one assumes that, although the density field has
undergone a substantial non--linear evolution, it is still
connected to the velocity by a linear relationship.
\end{description}

Non--linear effects in
the velocity field, such as infall and merging, are not accounted for
in this simulation method. These
effects, however, are only expected to dominate on scales $\mincir
10$--20$\hm$,
where the observational data on cluster peculiar velocities are in any
case rather unreliable.

\subsection{Comparison with N--body velocities}
In Paper I we noted that, while the ZA is very accurate at locating clusters in
the correct positions for models with $\sigma_8\mincir 1$ [$\sigma_{8}$ is the
rms fluctuation amplitude within a top--hat sphere of radius $8\hm$ radius],
non--linear gravitational
effects on the cluster mass scale degrade this precision when $\sigma_8\magcir
1$. However, this is not expected to represent a limitation for the present
analysis. In fact, in order to study the cluster velocity field it is not
necessary to know cluster positions with any great precision. What we need from
simulations is only a population of objects tracing the large--scale flows
having the same sampling density and same selection biases
as real clusters do.

Before presenting the results of our analyses, it is important to
assess the reliability of our approach to the simulation of the
cluster velocity field. We therefore begin by comparing
results of our approach  with
those obtained from the same N--body simulation we previously used in
Paper I for the
comparison of cluster positions. The initial spectrum corresponds to the
Cold+Hot DM (CHDM) model with 30\% of hot component (see next section). The
box size is $320\hm$ (with $h=0.5$) and the number of grid points and
particles is $128^3$. We do not distinguish between hot and cold particles,
since any effect of residual free-streaming should be negligible at the
smallest scales $r \magcir 2.5\hm$ allowed by our resolution.

In the N--body
simulation clusters, are identified as local maxima on the grid, following
the same method outlined in the previous subsection. The velocity at the
grid point positions is defined by following prescription {\bf (a)},
previously described, with the difference that non--linear (final)
particle velocities are now used and that density and momentum are smoothed
with a top--hat filter of width $10\hm$ to create continuous fields
and to ensure that the density is non--zero at each grid point.

For comparison we run a ZA simulation having the same initial spectrum and
random phase assignment and assuming the same box size and resolution. We
show results at two different evolutionary stages, corresponding to
$\sigma_8=0.67$ and $\sigma_8=1$. Even though
the two--year {\em COBE} data (Bennett
et al. 1994) are roughly consistent only with the first epoch, we consider
the second, higher, normalization as well, in order
to assess the reliability of the ZA
approach in a more evolved situation, when the cluster mass scale starts to
become non--linear. The filtering radii, $R_f$, applied in the
ZA simulation are $R_f=2.3$ and $R_f=4.5\hm$ for
$\sigma_8=0.67$ and $\sigma_8=1$,
respectively. In the following we will compare the velocity fields traced
by the clusters identified in the N--body simulation and in the ZA
simulation, by adopting the two definitions (LIN and ZEL) of velocity described
above.

\subsubsection{General features of the velocity fields}
In Figure 1 we compare the velocity fields traced by clusters
within a slice
60$\hm$ thick for the PM and ZA simulations. The fields are recovered
by smoothing the velocities associated
to the PM and ZA cluster positions  with a Gaussian window of radius $20\hm$.
The resulting velocity field is then reassigned
on $16^3$ grid points.
All clusters are equally weighted, independently of their masses.
At the first stage ($\sigma_8=0.67$), there is a strong correspondence between
the cluster velocity fields obtained by different methods: all the main
features are well
reproduced at the correct locations. At $\sigma_8=1$, although the principal
stream lines are still well delineated, there is a slight
tendency for both ZEL and LIN to
overestimate the velocity amplitude.

It is quite interesting to follow the evolution of the non--linear feature
at the centre of the panels displayed, where some degree of
infall (convergence of stream lines) takes place. At $\sigma_8=0.67$ this
feature is quite well reproduced by the LIN and ZEL velocities. However, at
$\sigma_8=1$ the higher degree of infall displayed by the N--body velocities
is not reproduced by the LIN and ZEL fields. As the degree of non--linearity
increases, shell--crossing becomes more and more important:
stream lines consequently cross each other, instead of simply converging
towards potential minima.

\subsubsection{Statistical tests}
In order to obtain a more quantitative assessment of the reliability of our
approach, we analyze PM, LIN and ZEL velocity fields with the same
statistical tests to be applied in the analysis of the data we
perform later. They are the 3D
velocity frequency distribution (VFD), the
velocity correlation function and
the bulk flow (as a function of scale).
Figure 2 summarizes the results at the two evolutionary
stages.

The VFD $P(v)$ is defined as the fraction of  cluster velocities
in a given range. The results of $P(v)$ for the 3D cluster velocity field
are shown in the upper panels. We used a Kolmogorov--Smirnov (KS) test to
compute the significance level for the null hypothesis that the different
VFDs are drawn from the same distribution. Comparing LIN to ZEL and PM,
we find significant departures at both  output times: the
distribution peak for LIN is always  shifted to lower
velocities ($\sim 300$ versus $\sim 450-500 \vel$ at $\sigma_8=0.67$ and
$\sim 500$ versus $\sim 600-650 \vel$ at $\sigma_8=1$). On the other hand,
the results for PM and ZEL are in good agreement: the KS test indicates
high probabilities both for both $\sigma_8=0.67$ and
$\sigma_8=1$.

Note that the VFD statistics do not take into account the
spatial distribution of clusters;
at the more evolved epoch the agreement of cluster positions
between PM and ZEL is not as good (see the discussion in Paper I).
In order to take into account the
cluster positions in the statistics of the velocity field, we will use
the velocity correlation tensor, defined as
\begin{equation}
\Psi_{ij}({\bf r})\equiv \langle v_i({\bf r_1}) v_j({\bf r_2}) \rangle.
\label{correlav}
\end{equation}
Here, ${\bf r_1}$ and ${\bf r_2}= {\bf r_1}+{\bf r}$  are the position vectors
of two objects having peculiar velocities $\bf v({\bf r_1})$ and $\bf v({\bf
r_2})$, respectively (G\'orski 1988; G\'orski et al. 1989). In
general what one estimates is the scalar velocity correlation function
$\xi_v(r)$ which is defined as the trace of the velocity correlation tensor,
$\xi_v= | \sum_i \Psi_{ii}|^{1/2}$. According to its definition, $\xi(v)$ is a
measure of the coherence of the velocity field: if at some scale a
coherent bulk flow
dominates the velocity field, $\xi_v$ will be positive;
infall or outflow velocities result in a negative contribution to $\xi_v$.

The velocity correlation function $\xi_v(r)$ is plotted in the central
panels of Figure 2. Error bars, shown only for the PM results, are
$1\sigma$ uncertainties computed using the bootstrap technique. The
agreement between different determinations is good at both epochs: only at
very small distances ($r\mincir 20 \hm$), some differences (always smaller
than 2$\sigma$) are present.

The last test we consider is the bulk velocity $V_{\rm bulk}$,
defined as the centre--of--mass velocity of a specified
region and given by the integral of the peculiar velocities ${\bf v(x)}$
over a selected volume specified by a selection function $\phi(\xvec)$:
\begin{equation}\label{eq:bulk}
{\bf V}_{\rm bulk}(r)=\int_{0}^{r} \phi(\xvec) \; {\bf v(x)} \;
{\rm d}\xvec .
\end{equation}
In our case the velocity field is traced by galaxy clusters and
eq.(\ref{eq:bulk}) becomes
\begin{equation}\label{eq:bulk2}
 {\bf V}_{\rm bulk}= \sum_i w_i {\bf v}_i/\sum_i w_i,
\end{equation}
where $i$ refers to different clusters inside the considered region. The
weights $w_i$ account for cluster masses, inhomogeneous sampling and
uncertainties in the cluster positions. The choice of the weighting scheme
is particularly important when one deals with observational biases, and
several different schemes have been proposed in literature. For the purpose
of this analysis, we are studying a perfectly sampled cluster
distribution, for which the cluster masses and peculiar velocities
are known, so the
natural weighting scheme to implement is simple number weighting
(i.e. $w_i=1$).

Results for the bulk velocity statistics are shown in the lower panels of
Figure 2. Also in
this case we plot the error bars only for the N--body results: they are
obtained as standard deviation of bulk flows computed for 10,000 different
observers. Note that the rather small box size is expected to result in a
significant underestimate of $V_{\rm bulk}$ at scales $\magcir 70$--100$\hm$.
However, here we are interested only in a relative comparison of the different
simulation methods rather than estimating the absolute value of the different
model bulk flows. At $\sigma_8=0.67$, ZEL
simulations provide a reliable estimate of $V_{\rm bulk}$ on scales larger than
$20
\hm$. The situation is even better at $\sigma_8=1$: here the agreement between
PM and ZEL is well inside the $1\sigma$ range at all scales. On the other hand,
the LIN simulation tends to underestimate the bulk flow at small scales when
$\sigma_8=1$.

We conclude from this analysis that the quasi--linear definition ZEL gives
a good representation of the velocity field traced by clusters over the
whole scale range, even at the stage $\sigma_8=1$, when substantial
non--linearity appears on the cluster mass scale. This is particularly
relevant for those statistics incorporating both
position and velocity information, such
as the velocity correlation function and bulk flow. We
will therefore use the ZEL velocity definition in our cluster
simulations from now on.

\subsection{Models for the power spectrum}
As we have explained, our aim is to use the cluster velocity field
to constrain
the value of the density parameter $\Omega_0$, both with and without a
cosmological constant term, $\Omega_{\Lambda}$, assuming the context
of a CDM scenario.
 We considered values of the density
parameters corresponding to
$\Omega_0=0.2,0.3,0.4,0.6,0.8$. In addition, we also
simulated two $\Omega_0=1$ models, namely the standard CDM model (SCDM) and
a Cold+Hot DM (CHDM) model with 30\% of hot component. We therefore end up
with 12 models for the forthcoming analysis.
They are listed in Table \ref{t:dm}, in which OCDM stands for the
open models and $\Lambda$CDM for the models with a non--zero
cosmological constant term.

For the CDM power spectra we assumed the expression provided by Efstathiou,
Bond \& White (1992), with the shape parameter $\Gamma = \Omega_0
h\exp(-\Omega_B-\Omega_B/\Omega_0)$, corrected according to the prescription
of Peacock \& Dodds (1994) to account for the presence of baryons (Sugiyama
1995). We take the density parameter in baryons to be
$\Omega_B=0.013h^2$, as suggested by standard nucleosynthesis
(e.g. Reeves 1994).
The transfer function provided by Holtzman (1989) is used for the CHDM
spectrum.
For each pair of ($\Omega_0,\Omega_\Lambda$) values, we
choose the Hubble parameter $h$ in such a way that the age of the universe
\be
t_0\,=\,H_0^{-1}\int_1^\infty {dx\over x \,\sqrt{\Omega_0 x^3+(1
-\Omega_0-\Omega_\Lambda)x^2+\Omega_\Lambda}}
\label{eq:t0}
\ee
(see, e.g., Peebles 1993) is $t_0 \simeq 12\,$Gyr for all the models,
except for SCDM and CHDM, for which we assume $h=0.5$, giving $t_0 \simeq
13\, $Gyrs. Although $t_0 \simeq 12$ Gyrs seems to be
slightly too small in comparison to
most current estimates of the age of the universe
(e.g. Chaboyer et al. 1995; arguments in favour
of $t_0\simeq 11$ Gyrs from
globular cluster ages have also been advanced, e.g., Shi 1995),
we adopt this value of $t_0$ for the low--$\Omega_0$
models so as to obtain almost the same age as
with the usual choice of $h=0.5$ for
the $\Omega_0=1$ models.

To normalize the two $\Omega_0=1$ models to COBE, we
use  $Q_{rms-PS}=20\mu K$ (G\'orski et al. 1994).
In the case of the OCDM models, we
resort to the normalization procedure outlined by G\'orski et al. (1995). The
procedure is straightforward to implement, since we select models with the same
criteria as G\'orski et al. (1995) did in their Table 1. We take for $\sigma_8$
the central value in their reported $1\sigma$ range, which is associated to
uncertainties in the $Q_{rms-PS}$ estimates.

For the $\Lambda$CDM models, Efstathiou et al. (1992) proposed the relation
\be
Q_{rms-PS}\,=\,\left({5\over 6\pi^2}\right)^{0.5}\left ({H_0\over 2c}\right)
^2T_0 \Omega_0^{0.77}B^{0.5}\,,
\label{eq:qla}
\ee
($T_0=2.735\,$K) which connects the purely Sachs--Wolfe (SW) quadrupole to
the amplitude $B$ of the primordial power spectrum, $P(k)=Bk$. However,
as, for example Stompor, G\'orski \& Banday (1995; hereafter SGB) have
pointed out,  significant
additional contributions to the radiation power spectrum come from the
integrated SW effect induced by the cosmological constant.
Unfortunately, these authors reported, in their Table 2, the
$\sigma_8$ normalizations only for models with two fixed values of $h$, namely
$h=0.5$ and $h=0.8$. In order to fix the normalization for the models
corresponding to different choices of $h$, we proceed as follows.
SGB noted that the integrated SW effect is negligible for $\Omega_0\magcir
0.5$. Therefore, we use for $\Omega_0 \ge 0.5$ the pure SW normalization
provided by eq.(\ref{eq:qla}). On the other hand, from Figure 3 of SGB, it
turns out that for our $\Omega_0 <0.5$ models the $\sigma_8$ normalization
essentially depends on the value of $\Omega_0 h^2$. Therefore, we compute
the $\Omega_0 h^2$ value for each model and use this plot to
normalize our spectra when $\Omega_0 <0.5$. We expect the accuracy of this
procedure to be within 10\%--15\%, of the same order of the statistical and
systematic deviations estimated by SGB in their analysis.

The values of $\sigma_8$ for all the considered models are shown in Table
\ref{t:dm}, where we also report the filtering radii $R_f$ used to suppress
shell--crossing in the ZA (see Section 2.1).

For each model we ran a single realization within a box of $960\hm$ on a
side,
using $256^3$ grid points and particles. With the adopted
cluster mean separation, we end up with about 16,000 objects in each
simulation box. The very large size of the box ensures
that: {\em (a)} no appreciable box--to--box variance is present in the final
results (we verified this by running two realizations for $\Lambda$CDM02 and
CHDM models);
{\em (b)} fluctuation modes with wavelengths larger than the box size
make a negligible contribution to bulk flows and velocity correlations.
Indeed, our analysis is confined within $180\hm$, which is roughly the
scale where the reconstructed bulk flow traced by Abell/ACO clusters
is reliable (Branchini, Plionis \& Sciama 1995; BPS hereafter).

\begin{table}[tp]
\centering
\caption[]{The model parameters.
Column 2: the density parameter $\Omega_0$; Column 3:
the cosmological constant term $\Omega_{\Lambda}$;
Column 4: the Hubble parameter $h$; Column 5: the linear rms
fluctuation amplitude at $8\hm$ $\sigma_8$; Column 6: the filtering radius
$R_f$ (in $\hm$).}
\tabcolsep 5pt
\begin{tabular}{lccccc} \\ \\ \hline \hline
 Model & $\Omega_0$ & $\Omega_{\Lambda}$ & $h$ & $\sigma_8$ & $R_f$
\\ \hline
SCDM & 1.0 & 0.0 & 0.50 & 1.36 & 6.0 \\
CHDM & 1.0 & 0.0 & 0.50 & 0.78 & 3.3 \\ \\

OCDM02 & 0.2 & 0.0 & 0.70 & 0.31 & 0.5 \\
OCDM03 & 0.3 & 0.0 & 0.65 & 0.50 & 1.6 \\
OCDM04 & 0.4 & 0.0 & 0.65 & 0.75 & 3.1 \\
OCDM06 & 0.6 & 0.0 & 0.60 & 1.10 & 4.9 \\
OCDM08 & 0.8 & 0.0 & 0.55 & 1.30 & 5.8 \\ \\

$\Lambda$CDM02 & 0.2 & 0.8 & 0.87 & 1.20 & 5.8 \\
$\Lambda$CDM03 & 0.3 & 0.7 & 0.78 & 1.44 & 6.9 \\
$\Lambda$CDM04 & 0.4 & 0.6 & 0.72 & 1.54 & 7.2 \\
$\Lambda$CDM06 & 0.6 & 0.4 & 0.64 & 1.47 & 6.6 \\
$\Lambda$CDM08 & 0.8 & 0.2 & 0.58 & 1.43 & 6.3 \\ \hline

\end{tabular}
\label{t:dm}
\end{table}

\section{Comparison between models}
In this section we consider the 3D velocity frequency distribution and
 the velocity correlation function, with the aim of
quantifying intrinsic differences between models. Any
comparison with observational data and how observational uncertainties are
taken into account in simulation analysis will be discussed in the next
section.

\subsection{The velocity frequency distribution}
As a first test, we consider the VFD $P(v)$ for the 3D cluster velocity field.
We compute the VFD for the whole box for each of the simulations.
In Figure 3 we plot the results for
SCDM and CHDM models (top left panel), the OCDM models (top right panel)
and $\Lambda$CDM models (bottom left panel). As expected, models with low
$\Omega_0$ generate smaller velocities, independently of the presence of
the cosmological constant. However, the $\Omega_\Lambda$ term tends to increase
the average peculiar velocity and is also effective in concealing
differences in the VFDs among various $\Omega_0$ choices.
Consequently, $\Lambda$CDM models
have VFDs that are similar to each other for $\Omega_0 \ge 0.4$, while the
analogous OCDM models show much better separated VFD curves. The model with
the largest velocities is SCDM while the probability of having very
large velocity in CHDM is smaller than in several of the low density models.

Due to their small size, we prefer not to plot the error bars, estimated as
scatter between the two realizations of CHDM and $\Lambda$CDM02 models.
Their smallness is a consequence of the large dimension of the simulation
box and confirms that the box size we adopted
is large enough to avoid any
significant
effect of cosmic variance. This is especially true for CHDM, which has a
smaller relative amount of large--scale power and, consistently, displays a
smaller dispersion than $\Lambda$CDM02. Therefore, we expect that
uncertainties in the analysis of realistic simulated samples (see next
section) will be associated with the limited number of included clusters and
to the observer--to--observer scatter, rather than to the cosmic variance
arising from the finite box size.

Considering N--body simulations of different cosmological scenarios (SCDM,
$\Lambda$CDM03 and $\Omega_0=0.3$ primeval baryonic isocurvature
models), Bahcall, Cen \& Gramann (1994) showed that the shape of the
cluster velocity distribution is well matched by a Maxwellian distribution
$P(v) \propto v^2 \exp (-v^2/2 \sigma_v^2)$, where $\sigma_v$ is the
velocity dispersion. This result is expected when a Gaussian distribution
for the initial density fluctuations is assumed and effects of non--linear
gravitational clustering are negligible at determining cluster velocities.
Using the value of $\sigma_v$ obtained from the VFDs of our simulations, we
found that a Maxwellian distribution provides a good representation for all
considered models. In Figure 3 the fitting curves are plotted (dashed
lines) for clarity only for the two $\Omega_0=1$ models: they are almost
indistinguishable from the simulated data and closely resemble
the result of Bahcall et al. (1994a), which is based on N--body
experiments.

\subsection{The velocity correlation function}
 Cen, Bahcall \&
Gramann (1994) showed that the velocity correlation function $\xi_v(r)$,
especially at small separation,
is strongly dependent on cosmological models,
and therefore represents a potentially useful tool for
 constraining them. However, with the observational data
currently available it is very difficult to obtain a determination of the
cluster velocity correlation function, which is sufficiently reliable
to be effectively
compared it with simulated data, although Croft \&
Efstathiou (1995) have recently succeeded in applying a maximum
likelihood analysis to the
velocity correlations estimated from
two sets of observational data, finding marginal evidence
for more power than the standard CDM model.

In Figure 4 we plot $\xi_v(r)$ for the considered models at
scales $r\ge 20\hm$, where the ZA description of the velocity correlations has
been shown to be reliable. Consistent with the results from the VFD analysis,
OCDM models are characterized by a progressively larger $\xi_v(r)$ as
$\Omega_0$ increases. A clear discrimination between models is
possible up to scales $r\simeq 60\hm$, while above $100\hm$ all the models seem
to converge, with the exception of OCDM02.
Among the $\Lambda$CDM models, the only
one with a significantly different behaviour is  $\Lambda$CDM02: all the
other spectra provide almost indistinguishable velocity correlations.

We note also that the velocity correlation functions of different
realizations for CHDM and $\Lambda$CDM02 models are extremely similar, thus
indicating that cosmic variance is negligible in  such big
simulation boxes.

\section{Comparing models and observations}
In this section we attempt to constrain the different cosmological
models by comparing predictions of properties of the cluster
peculiar velocity
field with relevant available data. To this end, we compute both the bulk
flow and the cumulative velocity frequency distribution. We prefer not to
consider velocity correlations, since estimating $\xi_v$ in a meaningful
way would require a large, statistically complete  sample of accurate cluster
peculiar velocities, which is not available at present.

\subsection{Cluster velocity data}
Unlike the simulation case, estimating the bulk flow executed by real
clusters is a non--trivial task, since the available cluster IRTF and
$D_{n}$--$\sigma$ peculiar velocity data are  sparse and
inhomogeneous. Furthermore, only the line--of--sight component of the
peculiar velocity is observed. These limitations can significantly affect
the determination of the observational bulk velocity (Kaiser 1988;
Reg\"{o}s \& Szalay 1989) and they have to be taken properly into account
if one wants to compare observational and model bulk flows.
Therefore, in the following, we will use the IRTF and $D_{n}$--$\sigma$
cluster peculiar velocities to compute only the cumulative velocity
frequency distribution (CVFD), $P(>v)$, i.e. the fraction of clusters
with  a velocity greater than $v$ (the VFD is its differential version) and
compare it with model predictions.

A possible way to avoid the above--mentioned problems in the context
of the bulk flow test is to reconstruct the full 3D cluster peculiar velocity
field from their observed redshift space distribution
[see Dekel (1994) and   Strauss \& Willick (1995) for reviews of the
 reconstruction methods available].
In the following analysis we will consider the
bulk velocity derived from the reconstructed Abell/ACO cluster density
field of Branchini \& Plionis (1995) and Branchini et al. (1995), hereafter
BPS. These
authors found a good consistency between their bulk flow and their derived 3D
smoothed density field, and independent measurements contained within
a depth of 8000 km s$^{-1}$ in the Mark III
galaxy sample
(cf. Dekel 1994; Hudson et al. 1995). We will also use the BPS velocities to
constrain model predictions of $P(>v)$.

\subsubsection{Observational velocities}
We construct our composite observational catalogue from cluster peculiar
velocity determinations by Aaronson et al. (1986), Mould et al.
(1991,1993), Han \& Mould (1992) and Mathewson, Ford \& Buchhorn (1992),
which are all based on IRTF measurements. We add the further
data reported in Table 7 by Mould et al. (1991), which contains peculiar
velocities obtained both with the IRTF method (Aaronson et al. 1989) and
with the $D_n$--$\sigma$ technique (Lucey \& Carter 1988; Faber et al. 1989).
Finally, we include cluster velocities obtained by Hudson (1994), both
with $D_n$--$\sigma$ and IRTF  techniques, and the recent IRTF data from
Giovanelli (1995; see also Giovanelli et al. 1995). Of course, it is possible
that more than one velocity determination is available for
any given cluster: in such cases
we have chosen to include in the catalogue only the
determination with the smallest error. The result of this selection
is an interim
sample of 65 objects, with a very inhomogeneous spatial distribution. Some
of these clusters show very large errors and we prefer to exclude them from our
analysis to obtain a more robust result. We therefore decided to
consider only objects with peculiar velocity uncertainties smaller than 850
km s$^{-1}$, excluding 2 objects, namely 2159--32 from Mathewson et al.
(1992) and Hercules from Aaronson et al. (1986). We have verified that our
conclusions do not change if we exclude the clusters with velocity errors
larger than 450 km s$^{-1}$, so as to end up with a smaller catalogue of 43
objects. Finally, we also reject the data of Cen45, whose determination is
uncertain and controversial, with different authors reporting greatly
discrepant results [i.e. $203\pm 383$ km s$^{-1}$ from Han \& Mould (1983),
while $1663\pm 336$ km s$^{-1}$ from Mould et al. (1991); note that both
results are obtained with the IRTF method]. At the end, the composite
sample contains 62 objects, whose sources are detailed in Table
\ref{t:catacomp}.

Due to the two different methods used (IRTF and $D_n$--$\sigma$ relations)
and to the absence of an inner calibration of the data coming from
different sources, our composite sample may be affected by some
basically uncontrollable biases. For these reasons, after discussing
results coming from the composite sample, we will also
present results obtained using data coming from
unique sources.

\begin{table}[tp]
\centering
\caption[]{The composite catalogue of observed cluster peculiar velocities.
Column 1: the source of the data; Column 2: the method used;
Column 3: the number of cluster peculiar velocities available in
the original paper; Column 4: the number of objects
actually used in the composite catalogue. }
\tabcolsep 5pt
\begin{tabular}{lccc} \\ \\ \hline \hline
 Source & Method & Available & Used  \\ \hline
Aaronson et al. (1986) & IRTF & $10$ & $0$ \\
Mould et al. (1991) $^a$& IRTF & $9$ & $6$ \\
Mould et al. (1991) $^b$& IRTF & $16$ & 1 \\
Mould et al. (1991) $^c$& $D_n$--$\sigma$  & $16$ & 1 \\
Han \& Mould (1992)& IRTF & $21$ & $6$ \\
Mathewson et al. (1992) &IRTF  & $24$ & $10$ \\
Mould et al. (1993)& IRTF & $8$ & $8$ \\
Hudson (1994) &IRTF & $21$ & $1$ \\
Hudson (1994) & $D_n$--$\sigma$ & $17$ & $6$ \\
Giovanelli (1995)& IRTF & $23$ & $23$ \\ \\
TOTAL  &   & & 62 \\ \hline
\end{tabular}
\label{t:catacomp}
{\leftline{$^a$ from Table 6.}}

{\leftline{$^b$ from Table 7 (Aaronson et al. 1989).}}

{\leftline{$^c$ from Table 7 (Lucey \& Carter 1988; Faber
et al. 1989).}}
\end{table}

\subsubsection{{\em BPS} reconstructed velocities}
Branchini et al. (1995) adopt a two step procedure  to recover  the real--space
positions and peculiar velocities
of Abell/ACO clusters, within $cz
\le 25,000$ km $s^{-1}$, starting from their redshift space positions.

\begin{itemize}
\item As a first step, BPS generate, via Monte Carlo techniques, several
different realizations of a synthetic cluster population, spatially
correlated with the observed clusters, so as to correct for selection effects
and properly homogenize the Abell and ACO catalogues. The outcome is an
all--sky volume limited distribution of synthetic + real clusters in
redshift space.
\item The second step is the application of
an iterative technique, similar to that
proposed by Strauss \& Davis (1988) and Yahil et al. (1991), which assumes
linear theory and linear biasing, to reconstruct the true cluster positions
from their observed redshifts.
\end{itemize}

The velocity of each cluster is, thus, essentially obtained from its
measured dipole evaluated at its reconstructed real space position.
Taking into account the uncertainties due to the intrinsic
error of their method as well as the uncertainties related to modelling the
observational parameters (galactic absorption, Abell/ACO
homogenization procedure, etc.), BPS estimated their mean cluster velocity
error to be $\approx 170$ km s$^{-1}$
with a dispersion of about 90 km s$^{-1}$ around this value.

We note that the BPS reconstruction method is based on two assumptions,
namely that: {\bf (a)} linear gravitational instability holds and
{\bf (b)} the simple linear biasing relation, $\delta n/n = b
(\delta \rho/\rho)$, connects fluctuations in cluster number counts and
DM density, with a position--independent biasing parameter $b$.
While the former assumption is
expected to be satisfied at the large scales relevant in our analysis, the
latter is probably an oversimplification. Although its effect,
in the light of the success of the reconstruction method
(cf. BPS), is likely not
to be dramatic  it could nevertheless introduce uncertainties into
the individual
cluster velocities which are not included in the previous quoted errors.

We will use the 280 clusters within $cz \le 20,000$ km s$^{-1}$, since
BPS consider their cluster velocity field to be reliable out to to this
depth. Beyond $cz=20,000$ km s$^{-1}$ the redshift selection function drops
exponentially while for $cz>25,000$ km s$^{-1}$ they have used
the rather crude assumption of a homogeneous universe in their
reconstruction algorithm.

\subsection{The Bulk Flow Analysis}
In order to perform a consistent comparison between simulations and real
data, one would desire in principle to reproduce the observational set--up
as closely as possible. We place observers at grid points and define their
local density contrast $\delta_{\rm obs}$ and  bulk velocity averaged over a
top--hat sphere of radius $R=7.5\hm$ centred on them. The local value of
$\delta_{\rm obs}$ is estimated by convolving the Fourier transform of the
fluctuation density field with the Fourier transform of the top--hat window,
$W(kR)=3(\sin kR-kR\cos kR)/(kR)^3$. The observer's velocity is defined
using the same procedure adopted for the cluster velocity. We estimate
$\delta_{\rm obs}$ and the peculiar velocity for 80,000 observers
chosen at random grid positions in each simulation. We then select those
`observers' that have the same characteristics as the Local Group, the
number of which is also considered as a test for DM models
(G\'orski et al. 1989; Tormen et al. 1993; Strauss et al. 1995; Moscardini
et al. 1995; Tini Brunozzi et al. 1995). These characteristics are:
\begin{description}
\item[(1)] peculiar velocity $V_{\rm LG}=627 \pm 44
\vel$ (error corresponding to $2\sigma$ uncertainties; Kogut et al. 1993)
for a top--hat sphere of radius $R=7.5\hm$ centred on the observer;
\item[(2)] density contrast within the same sphere in the range $-0.2\le
\delta_{LG} \le 1$.
\end{description}
We decided to neglect in this analysis a further requirement
concerning the quietness of the local flow, which implies a small value for
the local shear. In fact, we expect that any shear is poorly represented by
the velocity description provided by our simulations. In any case, as shown
by Moscardini et al. (1995), the local shear constraint is not a particularly
restrictive one.

In Column 2 of Table \ref{t:vlg} we report the fractions ${\cal F}_{LG}$ of
observers satisfying the LG requirements. The results for different
models are quite similar, with the highest probability occurring for OCDM06
and $\Lambda$CDM03 models. The only exceptions are represented by the open
models with $\Omega_0 \le 0.3$, where none of the chosen grid points
has the desired characteristics. This problem for the open models with
very low $\Omega_0$ and small $\sigma_8$ was already noted in Tormen et al.
(1993). Due to the impossibility of defining LG observers in these cases,
we prefer to consider in the following analysis 2000 observers randomly
chosen in the list of 80,000, dropping completely the LG constraints. We
checked that, at least for those models having a large enough number of
LG--like observers, this choice does not affect the results significantly.

A self--consistent comparison with the BPS bulk velocity would,
strictly speaking, require
us to apply their method also on the simulated cluster
distribution, starting from their redshift--space positions.
We have performed such a test on a subset of CHDM and
$\Lambda$CDM observers and we found a very good agreement
between the profiles of the ``true" and reconstructed bulk flows. On the
other hand, their amplitude can be matched a posteriori, by an
appropriate choice
of the (observer--averaged) cluster biasing parameter $b$
(we note that such definition of $b$ does not necessarily coincide with
that based on the ratio of rms fluctuations in cluster counts and DM
density; but see Kolokotronis et al. 1995).
Having assessed the agreement between ``true"
and reconstructued bulk flows, we decided to compare the BPS data to the
original ZEL cluster velocities, since applying the reconstruction
procedure for all the 2000 observers selected in each model would require
a very time--consuming computation.

The cluster bulk velocity for the simulations was computed as follows: for
each selected observer we measure the bulk velocity in spheres of
increasing radius using eq.(\ref{eq:bulk2}) with $w_i=1$ (we verified that
very similar results are obtained using a mass weighing scheme, i.e. with
$w_i \propto m_i$, where $m_i$ is the cluster mass). Due to the large
intercluster separation, we started measuring the bulk velocity within
spheres  of radius $50\hm$ to have reliable estimates. A limiting
depth of $180\hm$ has been chosen to allow a meaningful comparison with the
BPS data.

The scale--dependence of the bulk flow averaged over the 2000 observers is
plotted in Figure 5 for all the cosmological models we have considered. The
shaded region represents the 1$\sigma$ strip allowed by the BPS data. A
simple visual analysis reveals that the only models we can rule out at a
level larger than 1$\sigma$ are the OCDM ones with $\Omega_0 < 0.4$. Flat
models with non--vanishing cosmological constant, as well as CHDM and SCDM
are in good agreement with the data on scales larger than $\simeq 100\hm$.
On smaller scales only the CHDM model is well inside the observational
strip.
\begin{table}[tp]
\centering
\caption[]{
The fractions of
observers measuring a
particular characteristic
in different cosmological models. Column 2:
satisfying the Local Group requirements (see text; ${\cal F}_{LG}$);
Column 3: bulk flow
within 1$\sigma$ from the Branchini et al. (1995) value at
$50\hm$ (${\cal F}_{50}$); Column 4: bulk flow
within 1$\sigma$ from the Branchini et al. (1995)
value at $150\hm$ (${\cal F}_{150}$); Column
5: bulk flow
within 1$\sigma$ from the Lauer \& Postman (1994) value at $\sim100\hm$
(${\cal F}_{LP}$);
Column 6: bulk flow
within 2$\sigma$ from the Branchini et al. (1995) values
in the range from 50 to $150\hm$ (${\cal F}^{2\sigma}_{all}$).}
\tabcolsep 4pt

\begin{tabular}{lccccc} \\ \\ \hline \hline
Model & ${\cal{F}}_{LG}$ &  ${\cal{F}}_{50}$ &
${\cal{F}}_{150}$ &
${\cal{F}}_{LP}$ &
${\cal{F}}^{2\sigma}_{all}$  \\ \hline

SCDM & 0.026 & 0.293 & 0.365 & $0.032$ & 0.109  \\

CHDM & 0.053 & 0.437 & 0.315 & $0.019$ & 0.237  \\  \\

OCDM02 & $<10^{-3}$ & 0.017 & $< 10^{-3}$& $< 10^{-3}$ & 0.019  \\

OCDM03 & $<10^{-3}$ & 0.046 & 0.013 & $< 10^{-3}$ & 0.094 \\

OCDM04 & 0.015 & 0.383 & 0.108 & $< 10^{-3}$ & 0.228  \\

OCDM06 & 0.070 & 0.409 & 0.300 & $0.009$ & 0.187  \\

OCDM08 & 0.038 & 0.343 & 0.379 & $0.021$ & 0.151  \\ \\

$\Lambda$CDM02 & 0.052 & 0.396 & 0.385 & 0.047 & 0.278  \\

$\Lambda$CDM03 & 0.070 & 0.299 & 0.403 & 0.088 & 0.161  \\

$\Lambda$CDM04 & 0.051 & 0.274 & 0.418 & 0.068 & 0.141  \\

$\Lambda$CDM06 & 0.038 & 0.312 & 0.375 & 0.042 & 0.135  \\

$\Lambda$CDM08 & 0.030& 0.310 & 0.382 & 0.020 & 0.119  \\ \hline

\end{tabular}
\label{t:vlg}
\end{table}
As a more quantitative analysis, we estimate, for each model, the fraction of
observers whose measured bulk flow has certain specific properties. The
corresponding probabilities are also
reported in Table \ref{t:vlg}. In Column 3 we list the fraction of
observers ${\cal F}_{50}$ measuring a bulk flow at $50\hm$ within 1$\sigma$
from the BPS value. We choose this depth since it is still within the range
spanned by galaxy catalogues incorporating peculiar velocities. Moreover, at
$50\hm$ the BPS bulk velocity overlaps with that obtained from the POTENT
analysis (Dekel 1994) and with that measured using the recent da Costa et
al. (1995) data. Similarly, Column 4 contains the probability measured at
$150\hm$ (${\cal F}_{150}$). Although this is also the {\em limiting}
depth of the  Lauer \& Postman (1994) sample, we have compared their bulk
velocity
($V_{\rm bulk}\sim 700 \vel$), which is much higher than that of BPS,
at a depth of $\sim 100 \hm$ which corresponds to the {\em effective} depth
of their sample. The fraction of
observers measuring a bulk flow within 1$\sigma$ from the  Lauer \& Postman
(1994) one is listed in Column 5 (${\cal F}_{LP}$). Finally we measured the
fraction of observers whose cumulative bulk velocity is within the
2$\sigma$ strip of the BPS data throughout the range from $50$ to $150\hm$
(Column 6, ${\cal F}^{2\sigma}_{all}$). In this case we
preferred to consider a larger range (2$\sigma$) in order to have more
robust statistics. The comparison with data at $50\hm$ and $150\hm$ confirms
the visual impression: the only disfavoured models are the OCDM ones with
$\Omega_0=0.2 $ and $0.3$ and, more marginally and only at $150\hm$, the
$\Omega_0=0.4$ case. Moreover, for all models the Lauer \& Postman (1994)
observation is a very low probability event. The same conclusion was
reached using various techniques (Feldman
\& Watkins 1994; Tegmark, Bunn \& Hu 1994; Jaffe \& Kaiser 1994; Strauss et
al. 1995). Finally, in the open $\Omega_0=0.2 $ and $0.3$ CDM models,
a very tiny fraction of observers measures a cumulative bulk velocity in
accordance with the BPS data in the whole range of scales spanned by data;
the other models give similar results.

\subsection{Statistics of the velocity distribution}
In order to compare the data properly with the simulations,
we need to take account of
the (fairly large) uncertainties in cluster velocity determinations. To
this end, we convolve simulated cluster velocities using the following
procedure. Let $v_i$ be the intrinsic line--of--sight velocity for the
$i$--th model cluster ($i =1,\dots , 16000$) and $\delta v_j$ the velocity
error for the $j$--th randomly selected real cluster. The convolved
velocity, $\tilde{v}_i$, of the model cluster is obtained by randomly
generating it from a Gaussian distribution with mean $v_i$ and dispersion
$\delta v_j$. After repeating this operation for all the clusters in a
simulation, we end up with a sample of error--convolved cluster peculiar
velocities.

To estimate the uncertainty in the CVFD for the observational data, due to
individual velocity errors, we resample the real cluster velocities as
follows. For each cluster having raw velocity $v$ and error $\delta v$, we
generate a velocity $v'$ which is randomly taken from a Gaussian
distribution, having mean $v$ and dispersion $\delta v$. A resampling of
the original data set is therefore obtained by repeating this procedure for
all the clusters. We generate 20,000 resamplings of the observational
catalogues and for each of them we estimate the CVFD. The error
is estimated as the scatter within this ensemble.
The CVFD for the composite sample is plotted in Figure 6, where
for comparison we show also the CVFDs obtained considering
some of the single samples,
namely the Hudson (1994) IRTF sample, the Hudson (1994) $D_n$--$\sigma$ sample,
the Giovanelli (1995) sample and the Branchini et al. (1995)
reconstructed data (see the following subsections for more details about
these samples). For sake of
clarity we plot error bars, corresponding to the 2$\sigma$ scatter over the
observer ensemble, only for the Giovanelli (1995) and for the composite
samples.

\subsubsection{Results for the composite observational sample}
As a first characterization of the cluster velocity field, we compute for
our composite sample the
rms one--dimensional velocity, $\sigma_v\equiv \lb v_{1D}^2\rb^{1/2}$. The
corresponding values for the simulated catalogues are reported in Table
\ref{t:moment}, both before (Column 2) and after (Column 3)
error--convolution in the manner described above.
The latter have to be compared with results from the
line--of--sight observational peculiar velocities, which give
$\sigma_v=725\pm 60\vel$. The errors for the convolved $\sigma_v$ are
estimated as the 1$\sigma$ scatter over an ensemble of 20,000 samples, each
containing 62 randomly selected clusters taken from the whole simulation.
Note that the convolution with observational errors changes significantly
the values of $\sigma_v$. This is particularly true for the models with
small $\Omega_0$ (and small peculiar motions) where a large part of the
measured signal comes from the error noise. In Column 4 we report the
fraction ${\cal F}_{\sigma_v}$ of samples measuring values of the rms
velocity $\sigma_v$ within the 1$\sigma$ observational band. As our main
result, we note that observers in OCDM models with
$\Omega_0\le 0.4$ have a very low
probability of measuring the observed $\sigma_v$, while this value
 becomes typical if one considers $\Omega_0 \ge 0.6$. Among the
$\Lambda$CDM models, only the $\Omega_0= 0.2$ case is rather unlikely,
while all the other models are in good agreement with data. Both the
$\Omega_0=1$ models are consistent with the available observations.

\begin{table*}[table]
\centering
\caption[]{Results for the composite sample.
The different simulation rms
line--of--sight velocity values, $\sigma_v$ (in $\vel$), before
(Column 2) and after (Column 3) convolution with the observational
uncertainties of the composite sample.
Also reported is the fraction
${\cal F}_{\sigma_v}$ of the
extracted samples (Column 4) having $\sigma_v$ falling within the 1$\sigma$
observational range. Columns 5, 6 and 7 refer to the fractions of extracted
samples for which the measured cumulative velocity frequency distribution
$P(>v)$ assumes
a value within 1$\sigma$ from the observed data, computed for $v=0.5\sigma_v$
(${\cal F}_{0.5}$), $v=1\sigma_v$ (${\cal F}_{1}$) and $v=2\sigma_v$ (${\cal
F}_{2}$), respectively.}
\tabcolsep 5pt
\begin{tabular}{lcccccc} \\ \\ \hline \hline
 Model & $\sigma_v$ (unconv.) & $\sigma_v$ (conv.)&${\cal F}_{\sigma_v}$
 & ${\cal F}_{0.5}$ & ${\cal F}_{1}$ & ${\cal F}_{2}$ \\ \hline

 Composite ($D_{n}$--$\sigma$ \& IRTF) & & $725\pm 60$ & & & & \\

 SCDM & 664 & $797\pm 78$ & 0.413 & 0.187 & 0.133 & 0.543 \\
 CHDM & 467 & $633\pm 66$ & 0.293 & 0.548 & 0.505 & 0.276 \\ \\

 OCDM02 & 93 & $423\pm 61$ & $<10^{-3}$ & $<10^{-3}$ & 0.002 & 0.021 \\
 OCDM03 & 98 & $434\pm 62$ & 0.001 & $<10^{-3}$ & 0.002 & 0.035 \\
 OCDM04 & 197 &$465\pm 60$ & 0.002 & 0.013 & 0.009 & 0.039 \\
 OCDM06 & 415 &$592\pm 63$ & 0.123 & 0.503 & 0.379 & 0.183 \\
 OCDM08 & 574 &$718\pm 72$ & 0.594 & 0.373 & 0.390 & 0.507 \\ \\

 $\Lambda$CDM02 & 247 & $492\pm 59$ & 0.004 & 0.066 & 0.029 & 0.069 \\
 $\Lambda$CDM03 & 388 & $589\pm 63$ & 0.115 & 0.488 & 0.369 & 0.173 \\
 $\Lambda$CDM04 & 504 & $666\pm 68$ & 0.450 & 0.499 & 0.523 & 0.371 \\
 $\Lambda$CDM06 & 570 & $720\pm 72$ & 0.597 & 0.358 & 0.382 & 0.506 \\
 $\Lambda$CDM08 & 623 & $762\pm 73$ & 0.546 & 0.235 & 0.205 & 0.558 \\ \hline

\end{tabular}
\label{t:moment}
\end{table*}

The CVFDs resulting from both real data and simulations are
compared in Figure 7,
where the model curves represent the mean over 20,000 samples and the
observational error bars are 2$\sigma$ uncertainties. Again, models with
very low density are disfavoured, especially for the OCDM case:
they produce too few clusters with sufficiently high velocity. Moreover, we
note that, although the $\Lambda$CDM08 and SCDM models have the correct
$\sigma_v$, they nevertheless appear to have a systematically
smaller number of clusters with low velocities ($\mincir 600\vel$).
The other models
appear to lie within the 2$\sigma$ observational uncertainties for
the whole range of velocities considered.
The discriminatory power of this test is strongly reduced
by the smallness of the observational sample, as well as by  the size of the
uncertainties in the velocity measurements. In fact, as shown in Figure 2,
this test would be efficient in constraining the different models, were it
not for  the convolution with the observational errors which
has the effect of partially
washing out differences between the model CVFDs.

In order to perform a more quantitative analysis, for each model we
computed also the fraction of extracted samples for which the measured CVFD
assumes a value within 1$\sigma$ from the observed data. We are interested
in studying the velocity distribution for values corresponding to the
low--velocity tail, to the peak and to the high--velocity tail of the
distribution. Therefore, we decided to consider the CVFD at three different
values for the velocity, i.e. $v=0.5\sigma_v\equiv 362.5\vel$,
$v=1\sigma_v\equiv725\vel$ and $v=2\sigma_v\equiv1450\vel$, where we have
taken for $\sigma_v$ the observational value; we named the corresponding
fractions ${\cal F}_{0.5}$, ${\cal F}_{1}$ and ${\cal F}_{2}$ respectively.
For our data sample we find $P(>0.5\sigma_v)=0.537\pm 0.051$,
$P(>1\sigma_v)=0.245\pm 0.041$ and $P(>2\sigma_v)=0.065\pm 0.024$. The
results for the different models, also reported in Table \ref{t:moment},
confirm our previous qualitative analysis. The CVFD at low--velocity rules
out the OCDM models with $\Omega_0 \le 0.4$, and, more marginally,
$\Lambda$CDM08 and SCDM models; at high velocity the models in
disagreement with the real data are the open ones with $\Omega_0 \le
0.4$ again, and $\Lambda$CDM02.

\subsubsection{Results for single observational samples}
We consider now three separate samples of cluster peculiar
velocities, in order to study the stability of our previous results
with respect to the source and the method
used to obtain the data (i.e. IRTF vs. $D_n$--$\sigma$).
In particular we consider:
\begin{description}
\item[(1)] The Hudson (1994) $D_n$--$\sigma$ sample, reported in his Table 2.
This is a list of 17 objects with distance smaller than 8000 km s$^{-1}$.
Unlike the Burstein (1990) sample, from which these data originate,
Coma is used as the
calibrating cluster; the reported random error in the calibration of the
$D_n$--$\sigma$ relation is 3.7\%. As in the previous analysis, we prefer to
discard the data of Cen45 because of the large discrepancy with other
determinations.
\item[(2)] The Hudson (1994) IRTF sample, reported in his Table 3, a
list of 21 objects with a distance smaller than 8000 km s$^{-1}$, again
originating from Burstein (1990). The calibration is made using the nine
Aaronson et al. (1986) clusters which are located inside the region
considered. The random error in this calibration is 2\%.
\item[(3)] The Giovanelli (1995) sample, an updated version
of the Giovanelli et al. (1995) catalogue containing 23 cluster
peculiar velocities obtained by the IRTF method. The calibration comes
from those clusters which lie at a distance larger than 4000 km s$^{-1}$.
\end{description}

The results of the CVFD test for these samples have been already shown in
Figure 6. In Table \ref{t:subs} we report the values of the rms velocity
estimated from the line--of--sight observational peculiar velocities with the
same method previously applied to the composite sample. We note that, while the
results from the IRTF and $D_n$--$\sigma$ Hudson (1994) catalogues are fairly
consistent both with each other and with the value obtained from the
consolidated catalogue ($646\pm 102 \vel$ and $688\pm 82 \vel$ versus $725\pm
60 \vel$), the value of $\sigma_v$ from the Giovanelli (1995) sample is much
smaller, $356\pm 37 \vel$. The resulting discrepancy has a high confidence
level, ($\sim 3\sigma$) for both the $D_n$--$\sigma$ and the IRTF Hudson
subsamples.

\begin{table*}[table]
\centering
\caption[]{Results for single samples. The values of the simulation rms
velocities $\sigma_v$ (in $\vel$), evaluated for one--dimensional data
after convolution with the observational
uncertainties, and
the fraction ${\cal F}_{\sigma_v}$ of the
extracted samples having $\sigma_v$ falling within the 1$\sigma$
observational range are reported for the Hudson (1994) $D_n$--$\sigma$
catalogue (Columns 2 and 3), for the Hudson (1994) IRTF
catalogue (Columns 4 and 5), for the Giovanelli (1995) IRTF
catalogue (Columns 6 and 7) and for the Branchini et al. (1995)
reconstructed velocities (Columns 8 and 9).}
\tabcolsep 5pt
\begin{tabular}{lcccccccc} \\ \\ \hline \hline

Sample & \multicolumn{2}{c}{Hudson $D_n$--$\sigma$} &
\multicolumn{2}{c}{Hudson IRTF} &
\multicolumn{2}{c}{Giovanelli IRTF} &
\multicolumn{2}{c}{Branchini et al.} \\

  Objects  & \multicolumn{2}{c}{16$^*$}&\multicolumn{2}{c}{21}&
\multicolumn{2}{c}{23} & \multicolumn{2}{c}{280}
\\  \hline \hline

 Model & $\sigma_v$ &${\cal F}_{\sigma_v}$ &
$\sigma_v$ &${\cal F}_{\sigma_v}$ &
$\sigma_v$ &${\cal F}_{\sigma_v}$ &
$\sigma_v$ &${\cal F}_{\sigma_v}$
\\ \hline

 Observed & $646\pm 102$ & & $688\pm 82$ & & $356\pm 37$ & & $522\pm 15$ & \\
\\

 SCDM &  $786\pm 72$ & 0.303& $781\pm 73$& 0.343& $702\pm 64$& $<10^{-3}$
&  $703\pm 64$ & 0.003 \\
 CHDM &  $625\pm 61$ & 0.885& $614\pm 60$& 0.400& $513\pm 50$& 0.001
& $512\pm 50$ & 0.220 \\ \\

 OCDM02 &$407\pm 45$ & 0.003 & $395\pm 47$ &$<10^{-3}$ & $215\pm 27$ & 0.003
& $214\pm 27$ & $<10^{-3}$ \\
 OCDM03 &$418\pm 47$ & 0.008 & $405\pm 50$ &$<10^{-3}$ & $222\pm 27$ & 0.004
& $222\pm 28$ & $<10^{-3}$ \\
 OCDM04 &$454\pm 49$ & 0.042 & $440\pm 49$ &$<10^{-3}$ & $284\pm 29$ & 0.264
& $284\pm 30$ & $<10^{-3}$ \\
 OCDM06 &$581\pm 56$ & 0.735 & $572\pm 56$ &0.169 & $465\pm 43$ & 0.008
& $464\pm 43$ & 0.107 \\
 OCDM08 &$707\pm 68$ & 0.727 & $700\pm 67$ &0.659 & $616\pm 57$ & $<10^{-3}$
& $616\pm 57$ & 0.055 \\ \\

 $\Lambda$CDM02 &$482\pm 50$&0.111&$469\pm 50$&0.003&$331\pm 33$&0.672
& $330\pm 32$ & $<10^{-3}$ \\
 $\Lambda$CDM03 &$578\pm 55$&0.724&$570\pm 57$&0.163&$458\pm 43$&0.013
& $460\pm 44$ & 0.093 \\
 $\Lambda$CDM04 &$654\pm 62$&0.896&$646\pm 62$&0.575&$551\pm 50$&$<
10^{-3}$  & $551\pm 50$ & 0.199 \\
 $\Lambda$CDM06 &$709\pm 67$&0.718&$703\pm 68$&0.650&$613\pm 57$&$<10^{-3}$
& $612\pm 57$ & 0.059 \\
 $\Lambda$CDM08 &$751\pm 68$&0.483&$746\pm 69$&0.513&$668\pm 60$&$<10^{-3}$
& $667\pm 60$ & 0.010 \\
\hline

\end{tabular}
\label{t:subs}
{\leftline{\hspace{1.5truecm} $^*$ the data of Cen45 is discarded.}}
\end{table*}

In order to compare these results with the cosmological models, we convolve
the simulated cluster velocities with the observational uncertainties for
each sample separately: the resulting $\sigma_v$ are shown in Table
\ref{t:subs}. These values and their errors are the mean and the
1$\sigma$ scatter over an ensemble of 20,000 samples, respectively.
Also reported is the fraction ${\cal F}_{\sigma_v}$ of
those samples which measure a value of the rms velocity $\sigma_v$ within the
1$\sigma$ observational band. Once again, the results for the
$D_n$--$\sigma$ and the IRTF Hudson (1994) samples are very similar and in
good agreement with the results for the composite catalogue: the only
models we can reject on the basis of this analysis are  OCDM with
$\Omega_0 \le 0.4$ and $\Lambda$CDM02.

The situation is radically
different when the Giovanelli (1995) data are used. Due to the smaller
errors reported for this sample, the discriminatory power of the test is
very high and only two models survive: namely OCDM04 and $\Lambda$CDM02.
Note that these models are both rejected by either the composite or
the Hudson (1994) samples.

\subsubsection{Results for the BPS reconstructed velocities}

The results of the comparison between models and BPS cluster velocities are
also shown in Table \ref{t:subs}, where we report the BPS  rms line-of-sight
velocity $\sigma_v = 522\pm15$ km s$^{-1}$, estimated with the same
convolution method that was previously applied to the observational samples.
Following the same procedure as before, we find that the only models having
a non--negligible fraction  of ``good" observers are CHDM, $\Lambda$CDM04 and
OCDM06. It is interesting to note that now the SCDM model is rejected at a
high significance level, in agreement with the result based on the
Giovanelli (1995) data. Note that none of the models that satisfy
the BPS  velocities are the same time
accepted by the analysis based on the Giovanelli data, but
are accepted on the basis of the Hudson and composite samples.

It is worth recalling here that the high discriminatory power of such
reconstructed data is due to both their large number and
their small errors. However, as we already
mentioned, such errors include observational as well as intrinsic
uncertainties related to the reconstruction procedure but they do not account
for possible systematic effects related to the validity of the linear biasing
assumption on which the reconstruction method relies.
It is wise, therefore, to interpret these results only as
indications, but indications which are, nevertheless,
in agreement with the analysis of the
composite and Hudson data.

\section{Discussion}

This work was devoted to a thorough investigation of the velocity field as
traced by galaxy clusters. We have focussed our
attention on understanding whether such large--scale flows can be usefully
employed to constrain the values of the cosmological parameters
 $\Omega_0$ and  $\Omega_\Lambda$.

We have used a simulation technique based on the Zel'dovich
approximation, first testing this method carefully
by comparing its results with
a full N--body code. We have introduced
two different definitions of velocity in our simulations.
The first, linear, definition (LIN) amounts to the assumption that
the particle velocities are given by the gradient of the linear gravitational
potential estimated at the initial (Lagrangian) particle position. In the
second, quasi--linear, definition (ZEL) the particle velocity is estimated
through the gradient of the (Zel'dovich--evolved) potential evaluated
at the final (Eulerian) particle position. We compared the cluster velocity
fields corresponding to these two
prescriptions with that obtained from the
N--body simulation by applying different statistical tests, such the
velocity frequency distribution, the velocity correlation function and the
bulk flow. From this analysis we conclude that the quasi--linear definition
(ZEL) for the cluster velocities is consistent with N--body
results for scales $\magcir 10 - 20\hm$. This is true even for
high--normalization spectra, $\sigma_8\simeq 1$,  when a substantial
degree of non--linearity is present on the cluster mass scale.

In the framework of the CDM cosmogony, we considered models with $0.2 \le
\Omega_0\le 1$, both with flat ($\Lambda$CDM) and open (OCDM)
geometry. We also considered the $\Omega_0=1$ Cold+Hot DM (CHDM) model
with 30\% of hot component. For each model, we ran a single realization
within a large enough box, $960\hm$ on a side,
to render any effect of
cosmic variance entirely negligible.

We first compared predictions of the relevant models for both
the 3D cluster velocity frequency distribution (VFD), $P(v)$, and the
velocity correlation function, $\xi_v(r)$. These statistics have
been shown to possess an intrinsic discriminatory power between such
 models. This is particularly true for the OCDM models which
have rather different
VFDs and velocity correlations. The differences are instead smaller in the
case of the $\Lambda$CDM models, with the only exception of the
$\Omega_0=0.2$ case. We also found, in agreement with previous analyses
(e.g. Bahcall et al. 1994a), that all the models display a $P(v)$ which
is well fitted by a Maxwellian distribution.

In order to constrain the models we compared their predictions with
available cluster velocity data. As a first test we computed the
probability of having observers with the Local Group characteristics (i.e.
similar local velocity and density contrast).
The comparison for the bulk flow statistic
has been made with the results recently obtained
by Branchini et al. (1995), based on an iterative procedure used
to reconstruct the cluster real space positions and peculiar velocities.
Both analyses show that the only models which can be ruled out at a rather
high confidence level are the OCDM ones with $\Omega_0 < 0.4$. Note
that the bulk flow measured by Branchini et al. (1995) at $150\hm$ has a
much lower amplitude (and much better alignment with the CMB dipole) than
that obtained by Lauer \& Postman (1994), which however used
cluster peculiar velocities measured directly using a distance indicator
based on the properties of brightest cluster galaxies.
 We found, in agreement with other analyses, that the Lauer \& Postman
(1994) bulk flow is an extremely rare event in most of our
models, although in the case of
$\Lambda$CDM models (particularly for those with $\Omega_\Lambda =0.3$
and $0.4$) there is a small but non--negligible
probability of obtaining the sample amplitude as
the Lauer \& Postman bulk flow.

As a further test, we computed the 1D cumulative velocity frequency
distribution (CVFD) and the velocity dispersion $\sigma_v$. We
constructed a composite catalogue by collecting observed cluster velocities
from different data sets, based on both $D_n$--$\sigma$ and IRTF methods
(see Table \ref{t:catacomp}). The final observational sample used in the
analysis contains 62 objects. We took into account the large observational
uncertainties by convolving the simulated cluster velocities with the error
reported in the source catalogues. This convolution broadens the velocity
distributions and increases the values of $\sigma_v$, thus reducing the
differences between the models. In this way the possibility of constraining
the $\Omega_0$ and $\Omega_\Lambda$ parameters is largely reduced. The
low--$\Omega_0$ models are again disfavoured, particularly in the
case of open models, for which $\Omega_0 > 0.4$ is required to fit the
data; the $\Lambda$CDM02 model is also in trouble, while $\Lambda$CDM08 and
SCDM have systematically fewer clusters with high velocity than in the
observational sample.

In order to check whether our composite sample is affected by biases
associated with
its heterogenous composition, we considered also three smaller
samples directly taken from the literature: the $D_n$--$\sigma$ and the
IRTF Hudson (1994) catalogues and the IRTF Giovanelli (1995) catalogue. The
paucity of the data and the large uncertainties affecting the
two Hudson samples yield
less stringent results, but in agreement with the analysis of the composite
catalogue (cf. Tables \ref{t:moment} and \ref{t:subs}). Note that the two
Hudson samples contain data obtained from different distance indicators
($D_n$--$\sigma$ and IRTF relations). The similarity of the
results therefore
seems to indicate that there is not an intrinsic bias introduced by
the technique adopted in obtaining the data. On the other hand, the results
from the Giovanelli (1995) sample are completely different: this catalogue
is dominated by clusters with very small velocities and the
resulting velocity dispersion is consequently also
very small. In this case, the only
models in agreement with the data are OCDM04 and $\Lambda$CDM02, which were
excluded by the analysis of the composite and of the Hudson samples.

The discrepancy between the Giovanelli data and all other studies was
already noted by the authors themselves (see, e.g., Giovanelli et al.
1995). They claim that their results are more reliable for the following
reasons:
\begin{description}
\item[(1)] they used a homogenous galaxy sample, which provides a full
coverage of the sky;
\item[(2)] they adopted a different Tully--Fisher template relation,
obtained from an extensive study of clusters;
\item[(3)] they applied an internal extinction correction which depends on
the luminosity and allows for larger flux corrections.
\end{description}

On the other hand, a possible bias for the Giovanelli sample might have
arisen from their estimate of the IRTF relation zero--point.
They adopted that of the mean
relation obtained from all clusters with $cz > 4000$ km s$^{-1}$ in the CMB
frame, since, as they argue, at such distances small peculiar velocities
translate into small magnitude offsets, and that possible bulk flows do not
affect the mean IRTF relation since the clusters are isotropically
distributed. However, since the number of available clusters is small
($\sim 15$), even if they are random samplers of an isotropic distribution,
they could introduce shot--noise deviations from the `isotropic'
zero--point value. A second concern is related to the use of the CMB
rest--frame. If there is a bulk flow, in which the LG participates, then, by
using this frame the sampled volume becomes elongated along the
direction of the bulk flow, compared to a spherically
symmetric distribution.

We have also analysed the CVFD for the
BPS reconstructed cluster velocities, which represents a completely
independent data set. It turns out that such velocities pose the most
stringent constraints on the different cosmological models (cf. Table
\ref{t:subs}). Since, however, the BPS velocities are not truly observational
but depend on various assumptions we choose not to present the above results
as final and firm conclusions. In particular, the assumption of a linear
biasing relationship which is independent of spatial position,
may well not be valid. Our ignorance about
the possible cluster--by--cluster variation of the biasing parameter would
increase the individual reconstructed velocity uncertainties. For these
reasons, we prefer at present to consider these results only as
circumstantial evidence in support of inferences made on the basis
of the composite and Hudson data.

\section{Conclusions}

The first definite conclusion of this work is  that our method to
simulate the cluster velocity field, which is based on an optimized version
of the Zel'dovich approximation, is easy to implement, cheap from the
computational point of view and fully reliable, as shown by the comparison
with the results of a full N--body code.

As for the comparison between observational data and numerical simulations
on cluster peculiar velocities is concerned,
our main findings can be sketched as follows.
\begin{description}
\item[(a)]
The bulk flow analysis performed on the Branchini et al. (1995) data
and the probability of reproducing the Local Group characteristics
rule out only open CDM models with $\Omega_0 < 0.4$.
We also confirm that the Lauer \& Postman (1994) high bulk velocity,
at an effective depth of $\sim 100 \hm$, is an extremely unlikely
event in all the $\Omega_\Lambda= 0$ models. However, in
$\Omega_\Lambda \ne 0$ models there
is a small but non--negligible probability of reproducing this observation.
\item[(b)]
The VFD analysis based on a composite sample of 62 cluster velocities,
obtained both with IRTF and $D_n$--$\sigma$ methods, rules out open CDM
models with $\Omega_0 \le 0.4$ and the $\Lambda$CDM model with
$\Omega_0=0.2$. Standard CDM and $\Lambda$CDM with $\Omega_0 =0.8$ are
disfavoured, although only marginally.
Similar results are found by applying the same test on the two
Hudson (1994) catalogues, based on the IRTF and $D_n$--$\sigma$ relations,
respectively: no systematic
bias appears to be introduced by the different distance
indicators.
\item[(c)]
The application of the VFD test to the IRTF Giovanelli (1995) sample gives
completely different results. The smaller velocities and
errors reported in this
sample are such that the only surviving models are the open CDM with
$\Omega_0 =0.4$ and the $\Lambda$CDM with $\Omega_0 =0.2$.
\item[(d)]
We found also that
the above two models, which are consistent with the Giovanelli data, are
strongly disfavoured by
an analysis of the Branchini et al. (1995) reconstructed velocities,
in agreement with the interpretation of the Hudson and composite
sample analyses.
\end{description}

As a concluding remark, we would like to stress the importance
of identifying the reasons for
the puzzling discrepancy between  different data sets.
Our analysis has shown that a small, but accurate, sample of cluster
peculiar velocities, like that of Giovanelli (1995), can put rather stringent
constraints on different DM models. We therefore consider
the analysis presented here as proving the potential of cluster
velocity considerations, and look forward to the
availability of newer and more precise cluster velocity measurements.

\section*{Acknowledgments}
LM and PTB thank Italian MURST for partial financial support. PTB also
acknowledges CUC for partial financial support. MP has been supported by an
EEC {\em Human Capital and Mobility} fellowship. PC is a PPARC Advanced
Research Fellow. EB and PTB acknowledge the EC for financial support during
their visit in QMWC and thank QMWC for its hospitality.
This work has been partly supported by funds originating
from the {\em EC} Human Capital and Mobility Network (Contract Number
CHRX--CT93--0129). We thank Riccardo Giovanelli for providing us with his
cluster redshifts prior to publication and for useful comments.

\newpage
\section*{Figure Captions}

\noindent
{\bf Figure 1.}
Comparison of the projected  velocity fields traced by clusters
in a N--body simulation (PM, upper
row) and in Zel'dovich simulations with two different definitions
for the velocity: ZEL (central row) and LIN (bottom row).
The simulations have a box--size of $320\hm$ and the same
CHDM initial conditions.
The slice is $60\hm$ thick.
A smoothing by a Gaussian filter with width $20\hm$ is applied.
Two different stages are shown: $\sigma_8=0.67$ (left column) and
$\sigma_8=1$ (right column).

\noindent
{\bf Figure 2.}
Comparison of the results of the application of different statistical
tests to the cluster velocity fields obtained
in a N--body simulation (PM, solid lines)
and in Zel'dovich simulations with two different definitions
for the velocity: ZEL (dotted lines) and LIN (dashed lines).
Upper row: the 3D velocity frequency distribution $P(v)$.
Central row: the velocity correlation function $\xi_v$ (in km s$^{-1}$)
as a function of the distance $r$ (in $\hm$).
Bottom row: the bulk flow velocity $V_{\rm bulk}$ (in km s$^{-1}$)
as a function of the distance $r$ (in $\hm$).
Two different stages are shown: $\sigma_8=0.67$ (left column) and
$\sigma_8=1$ (right column).

\noindent
{\bf Figure 3.} The frequency distribution $P(v)$ for the 3D cluster velocity
field for different models: SCDM and CHDM models (top left), OCDM models (top
right), $\Lambda$CDM models (bottom left).
The dashed lines in the top left panel
show the Maxwellian curves fitting the distributions.

\noindent
{\bf Figure 4.}
The velocity correlation function $\xi_v$ (in km s$^{-1}$)
as a function of the distance $r$ (in $\hm$) for different models:
SCDM and CHDM models (top left), OCDM models (top right),
$\Lambda$CDM models (bottom left).

\noindent
{\bf Figure 5.} The bulk flow velocity $V_{\rm bulk}$ (in km s$^{-1}$)
as a function of the distance $r$ (in $\hm$) for different models:
SCDM and CHDM models (top left), OCDM models (top right),
$\Lambda$CDM models (bottom left).
The dashed region represents the 1$\sigma$ strip allowed by the
Branchini et al. (1995) data.

\noindent
{\bf Figure 6.}
The 1D cumulative velocity frequency distribution $P(>v)$ of our composite
catalogue (solid line) is compared to that
of the Hudson (1994) IRTF sample (short--dashed line),
of the Hudson (1994) $D_n$--$\sigma$ sample (long--dashed line),
of the Giovanelli (1995) sample (dotted line) and of the Branchini et al.
(1995) reconstructed data
(dotted--dashed line).
Error bars, plotted only for the composite and Giovanelli samples,
correspond to the $2\sigma$ scatter over 20,000 resamplings.

\noindent
{\bf Figure 7.}
The 1D cumulative velocity frequency distribution $P(>v)$ for different models
compared with that resulting from our composite observational catalogue
(filled circles with
2$\sigma$ uncertainties
obtained as scatter over an ensemble consisting
of 20,000 resamplings of the
original data). The model curves are the mean of 20,000 samples of 62
randomly selected clusters whose velocities are
convolved with the observational
errors. The panels refer to
SCDM and CHDM models (top left), OCDM models (top right)
and $\Lambda$CDM models (bottom left).


\vspace{1.5cm}

\begin{thebibliography}{}
\bibitem[]{}Aaronson M., Bothun G., Mould J., Huchra J., Schommer R.A.,
Cornell M.E., 1986, ApJ, 302, 536
\bibitem[]{}Aaronson M., et al., 1989, ApJ, 338, 654
\bibitem[]{}Bahcall N.A., Cen R., Gramann M., 1994a, ApJ, 430, L13
\bibitem[]{}Bahcall N.A., Gramann M., Cen R., 1994b, ApJ, 436, 23
\bibitem[]{}Bennett C.L., et al., 1994, ApJ, 436, 423
\bibitem[]{}Borgani S., Plionis M., Coles P., Moscardini L., 1995, MNRAS,
in press, preprint ASTRO--PH/9505050 (Paper I)
\bibitem[]{}Branchini E.,  Plionis M., 1995, ApJ, in press
\bibitem[]{}Branchini E., Plionis M., Sciama D.W., 1995, ApJL, submitted [BPS]
\bibitem[]{}Bucher M., Goldhaber A.S., Turok N., 1995,
Phys. Rev. D., 53, 3314
\bibitem[]{}Burstein D., 1990, Mark II compilation of
peculiar velocities, privately circulated
\bibitem[]{}Cen R., Bahcall N.A., Gramann M., 1994, ApJ, 437, L51
\bibitem[]{}Chaboyer B., Kernan P.J., Krauss L.M., Demarque P., 1995,
preprint ASTRO-PH/9509115
\bibitem[]{}Coles P., Ellis G.F.R., 1994, Nature, 370, 609
\bibitem[]{}Coles P., Melott A.L., Shandarin S.F., 1993, MNRAS, 260, 765
\bibitem[]{}Croft R.A.C., Efstathiou G., 1994, MNRAS, 268, L23
\bibitem[]{}Croft R.A.C., Efstathiou G., 1995, in eds. Muecket J., et al.,
Proc. of the Potsdam Meeting on Large--scale structure in the universe,
in press
\bibitem[]{}da Costa et al., 1995, in
eds. Balkowski C., Maurogordato S., Tao C. \& Tr\^an Thanh V\^an J.,
Proc. of the Moriond Astrophysics Meeting on Clustering in the Universe,
in press
\bibitem[]{}Dekel A., 1994, ARA\&A, 32, 99
\bibitem[]{}Efstathiou G., Bond J.R., White S.D.M., 1992, MNRAS, 258, 1p
\bibitem[]{}Efstathiou G., Sutherland W.J., Maddox S.J., 1990, Nat, 348, 705
\bibitem[]{}Ellis G.F.R., Lyth D.H., Mijic M.B., 1991, Phys. Lett., B271, 52
\bibitem[]{}Faber S.M., Wegner G., Burstein D., Davies R., Dressler A.,
Lynden--Bell D., Terlevich R., 1989, ApJSS, 69, 763
\bibitem[]{}Feldman H.A., Watkins R., 1994, ApJ, 430, L17
\bibitem[]{}Giovanelli R., 1995, in
eds. Bonometto S., Primack J., \& Provenzale A.,
Proc. of the International School of
Physics Enrico Fermi, Course CXXXII: Dark Matter in the Universe, in press
\bibitem[]{}Giovanelli R., Haynes M.P., Chamaraux P., da Costa L.N., Freudling
W., Salzer J.J., Wegner G., 1995, in ed. Kafatos M.,
Proc. of the IAU Symp. Examining the
Big Bang and Diffuse Background Radiation, in press
\bibitem[]{}G\'orski K., 1988, ApJ, 332, L7
\bibitem[]{}G\'orski K.M., Ratra B., Sugiyama N., Banday A.J., 1995,
ApJ, 444, L65
\bibitem[]{}G\'orski K., et al., 1989, ApJ, 344, 1
\bibitem[]{}G\'orski K., et al., 1994, ApJ, 430, L89
\bibitem[]{}Gramann M., Bahcall N.A., Cen R., Gott J.R., 1995, ApJ, 441, 449
\bibitem[]{}Haenhelt M.G., Tegmark M., 1995, preprint ASTRO--PH/9507077
\bibitem[]{}Han M., Mould J.R., 1992, ApJ, 396, 453
\bibitem[]{}Hockney R.W., Eastwood J.W., 1981, Computer Simulations Using
Particles. McGraw--Hill, New York
\bibitem[]{}Holtzman J.A., 1989, ApJS, 71, 1
\bibitem[]{}Hudson M.J., 1994, MNRAS, 266, 475
\bibitem[]{}Hudson M.J., Dekel A., Courteau S., Faber S.M., Willick J.A.,
1995, MNRAS, 274, 305
\bibitem[]{}Jaffe A., Kaiser N., 1994, preprint ASTRO--PH/9408046
\bibitem[]{}Kaiser N., 1988, MNRAS, 239, 149
\bibitem[]{}Kofman L., Bertschinger E., Gelb J.M., Nusser A., Dekel A., 1994,
ApJ, 420, 44
\bibitem[]{}Kofman L., Gnedin N.Y., Bahcall N.A., 1993,  ApJ, 413, 1
\bibitem[]{}Kogut A., et al., 1993, ApJ, 419, 1
\bibitem[]{}Kolokotronis V., Plionis M., Coles P., Borgani S., Moscardini
L., 1995, MNRAS, submitted
\bibitem[]{}Lauer T., Postman M., 1994, ApJ, 425, 418
\bibitem[]{}Liddle A.R., Lyth D.H., Roberts D., Viana P.T.P., 1995,
preprint ASTRO--PH/9506091
\bibitem[]{}Linde A., Mezhlumian A., 1995, preprint ASTRO--PH/9506017
\bibitem[]{}Lucchin F., Matarrese S., 1985, Phys. Lett., 164B, 282
\bibitem[]{}Lucey J., Carter D., 1988, MNRAS, 235, 1177
\bibitem[]{}Mathewson D.S., Ford V.L., Buchhorn M., 1992, ApJSS, 81, 413
\bibitem[]{}Melott A.L., Pellman T.F., Shandarin S.F., 1994, MNRAS, 269, 626
\bibitem[]{}Moscardini L., Tormen G., Matarrese S., Lucchin F., 1995, ApJ,
442, 469
\bibitem[]{}Mould J.R., Akeson R.L., Bothun G.D., Han M., Huchra J.P.,
Roth J., Schommer R.A., 1993, ApJ, 409, 14
\bibitem[]{}Mould J.R., et al., 1991, ApJ, 383, 467
\bibitem[]{}Peacock J.A., Dodds S.J., 1994, MNRAS, 267, 1020
\bibitem[]{}Peebles P.J.E., 1993, Principles of Physical Cosmology.
Princeton University Press, Princeton
\bibitem[]{}Plionis M., Borgani S., Moscardini L., Coles P., 1995, ApJ, 441,
L57
\bibitem[]{}Primack J.R., 1995, Plenary Talk at Snowmass 1994, preprint
ASTRO--PH/9503020
\bibitem[]{}Ratra B., Peebles P.J.E., 1994, ApJ, 432, L5
\bibitem[]{}Ratra B., Peebles P.J.E., 1995, Phys. Rev., D52, 1837
\bibitem[]{}Reeves H., 1994, Rev. Mod. Phys., 66, 193
\bibitem[]{}Reg\H{o}s E., Szalay A.S., 1989, ApJ, 345, 627
\bibitem[]{}Riess A., Press W., Kirshner R., 1995, ApJ, 438, L17
\bibitem[]{}Sasaki M., Tanaka T., Yamamoto K., Yokoyama J.,
1993, Phys. Lett., B317, 510
\bibitem[]{}Shandarin S.F., Zel'dovich Ya.B., 1989, Rev. Mod. Phys., 61,
185
\bibitem[]{}Shi X., 1995, ApJ, 446, 637
\bibitem[]{}Stompor R., G\'orski K.M., Banday A.J., 1995, MNRAS, in press,
preprint ASTRO--PH/9502035 (SGB)
\bibitem[]{}Strauss M.A., Cen R., Ostriker J.P., Lauer T.R., Postman M.,
1995, ApJ, 446, 507
\bibitem[]{}Strauss M.A., Davis, M., 1988,
in Rubin V.C., Coyne G., eds, Large Scale Motions in the
Universe: A Vatican Study Week. Princeton Univ. Press, Princeton, p. 255
\bibitem[]{}Strauss M.A., Willick J.A., 1995, Phys Rep., 261, 271
\bibitem[]{}Sugiyama N., 1995, preprint ASTRO--PH/9503089
\bibitem[]{}Sunyaev R., Zel'dovich Ya.B., 1980, MNRAS, 190, 413
\bibitem[]{}Tegmark M., Bunn E.F., Hu W., 1994, ApJ, 434, 1
\bibitem[]{}Tini Brunozzi P., Borgani S., Plionis M., Moscardini L.,
Coles P., 1995, MNRAS, in press, preprint ASTRO--PH/9506110
\bibitem[]{}Tormen G., Moscardini L., Lucchin F., Matarrese S., 1993, ApJ,
411, 16
\bibitem[]{}Yahil A., Strauss M.A., Davis M., Huchra J.P., 1991, ApJ, 372, 380
\bibitem[]{}Yamamoto K., Bunn E.F., 1995, preprint ASTRO--PH/9508090
\bibitem[]{}Zel'dovich Ya.B., 1970, A\&A, 5, 84
\end{thebibliography}
\end{document}